\documentclass[preprint,12pt]{elsarticle}

\usepackage{amssymb}
\usepackage{amsmath}
\journal{Computational Materials Science}

\begin{document}

\begin{frontmatter}



\title{Influence of Local Icosahedral Short-Range Order on the Magnetization Dynamics of Amorphous Cobalt-Iron Nanodisks} %% Article title


\author[uchile]{Matias Sepulveda--Macias} %% Author name
            
\author[uchile]{Erick--Burgos Parra} %% Author name

\affiliation[uchile]{organization={Departamento de Fisica, Facultad de Ciencias, Universidad de Chile },%Department and Organization
          addressline={Las Palmeras 3425, Ñuñoa}, 
            city={Santiago},
            country={Chile}}
%% Abstract
\begin{abstract}
The microscopic origin of soft magnetic properties in amorphous alloys is fundamentally linked to the interplay between local topological disorder and magnetic exchange interactions. In this work, we employ a multiscale Spin-Lattice Dynamics (SLD) approach to investigate the magnetostructural correlations in amorphous Co$_{x}$Fe$_{1-x}$ nanodisks ($x=35, 50, 65$). By integrating classical molecular dynamics with a generalized magnetic Hamiltonian, we capture the dynamic feedback loop between lattice vibrations and spin precession. Topological analysis via Voronoi tessellation reveals a persistent species-dependent structural heterogeneity: Cobalt atoms preferentially adopt "solid-like" icosahedral packing, forming a rigid structural backbone, whereas Iron atoms exhibit a higher propensity for "liquid-like" disordered environments. We demonstrate that this topological disparity dictates the macroscopic magnetic response. The Cobalt-driven structural stiffness preserves a robust exchange network that maximizes saturation magnetization, while the local disorder inherent to Iron-rich regions introduces exchange fluctuations that act as an intrinsic damping mechanism, delaying magnetic relaxation. These findings provide an atomistic explanation for the stability of ferromagnetic order in Co-Fe metallic glasses and offer a pathway for tuning damping parameters in amorphous spintronic devices through stoichiometric control.
\end{abstract}



%% Keywords
%\begin{keyword}



%\end{keyword}

\end{frontmatter}



\section{Introduction}
\label{sec1}

Amorphous ferromagnetic alloys, particularly those based on the Cobalt-Iron (Co-Fe) system, have attracted significant attention in the field of condensed matter physics and spintronics due to their exceptional soft magnetic properties, including high saturation magnetization, low coercivity, and high permeability \cite{McHenry1999, Lubrorsky1980, Bai2018}. Unlike their crystalline counterparts, these metallic glasses lack long-range translational order; however, they exhibit a pronounced short-to-medium range order (SRO) that fundamentally dictates their mechanical and magnetic stability \cite{Sheng2006, Miracle2004}. Understanding the atomistic origin of this stability is crucial, as local structural heterogeneities—often described as ``solid-like'' clusters versus ``liquid-like'' disordered regions \cite{Egami1984, Cheng2011}—can dramatically influence spin-wave propagation, magnetic anisotropy, and relaxation processes \cite{Tsvelick01011983, BHATTI2017530, KAUL19855}.

The theoretical description of magnetic excitations in these disordered systems presents a complex multiscale challenge. Conventional micromagnetic simulations typically treat the material as a continuous medium, averaging out the atomic-scale fluctuations that are intrinsic to the amorphous state \cite{Evans2014}. On the other hand, purely classical Molecular Dynamics (MD) captures the structural evolution but neglects the magnetic degrees of freedom. To fully resolve the interplay between the disordered lattice and the magnetic system, it is necessary to treat atomic positions and magnetic spins equivalently \cite{Ma2011}. The Spin-Lattice Dynamics (SLD) framework addresses this by simultaneously solving lattice equations and the Landau-Lifshitz-Gilbert (LLG) equation \cite{Ma2012, Beaujouan2012}.  Implemented via the \texttt{SPIN} package in LAMMPS \cite{Tranchida2018, lammps}, this massively parallel, symplectic approach captures the critical feedback loop between phonons and magnons, which is essential for systems with significant magnetostructural coupling \cite{Tranchida2018, Ma2012}.

In this work, we employ this multiscale computational approach to investigate the magnetic dynamics of amorphous Co$_{x}$Fe$_{1-x}$ nanodisks. While experimental realization of pure amorphous Co-Fe alloys typically requires metalloid stabilizers (e.g., CoFeB, CoFeSiB) \cite{Oogane2006}, our study models the pure binary metallic glass to isolate the intrinsic influence of the transition metal topology on the exchange stiffness and magnetic relaxation, unperturbed by non-magnetic impurities.. We focus on the correlation between local topological motifs—characterized via Voronoi tessellation \cite{Stukowski2010}—and the macroscopic magnetic response across different stoichiometries (Co$_{50}$Fe$_{50}$, Co$_{65}$Fe$_{35}$, and Co$_{35}$Fe$_{65}$). By mapping the local atomic environments to a distance-dependent exchange interaction model, we demonstrate how the structural ``backbone'' of Cobalt-centered icosahedra stabilizes the ferromagnetic order, providing new insights into the intrinsic damping mechanisms and saturation limits of amorphous magnetic nanostructures.

\section{Computational Modelling}

\subsection{Theoretical Framework}
In this study, we employ a multiscale computational approach that integrates classical molecular dynamics (MD) with atomistic spin dynamics (SD). These simulations were performed using the LAMMPS software package~\cite{lammps} in conjunction with the SPIN package developed by Tranchida {\it et al.}~\cite{Tranchida2018}. The dynamics of the system are governed by a generalized Hamiltonian that couples the lattice and magnetic degrees of freedom:

\begin{equation}
\mathcal{H} = \sum_{i=1}^N \frac{\left|p_i\right|^2}{2m_i} + \sum_{i,j,i\neq j}^N V(r_{ij}) + \mathcal{H}_{\text{mag}} 
\end{equation}

where the first two terms represent the kinetic energy and the interatomic potential energy of the lattice, respectively. The final term, $\mathcal{H}_{\text{mag}}$, represents the magnetic contribution to the total energy, defined as:

\begin{equation}
\mathcal{H}_{\text{mag}} = -\sum_{i,j}^N J_{ij}(r_{ij})~\vec{s}_i\cdot\vec{s}_j - g\sum_{i=0}^N \mu_i\,\vec{s}_i\cdot \vec{B}_{\text{ext}} -\sum_{i=0}^N K_{\text{an}}(r_i)\left(\vec{s}_i\cdot \vec{n}_i\right)^2 
\end{equation}

The magnetic Hamiltonian includes a Heisenberg term for spin-spin interactions, where $\vec{s}_i$ is the normalized spin vector of atom $i$ and $J_{ij}(r_{ij})$ is the distance-dependent exchange coupling constant. The second term describes the Zeeman interaction with an external magnetic field $\vec{B}_{\text{ext}}$, where $g$ is the Landé factor and $\mu_i$ is the atomic magnetic moment. The third term accounts for the magnetocrystalline anisotropy, where $K_{\text{an}}$ and $\vec{n}_i$ define the anisotropy intensity and direction, respectively.

\subsection{Sample Preparation and Structural Characterization}

\begin{figure}[t]
\centering
\includegraphics[width=1.0\textwidth]{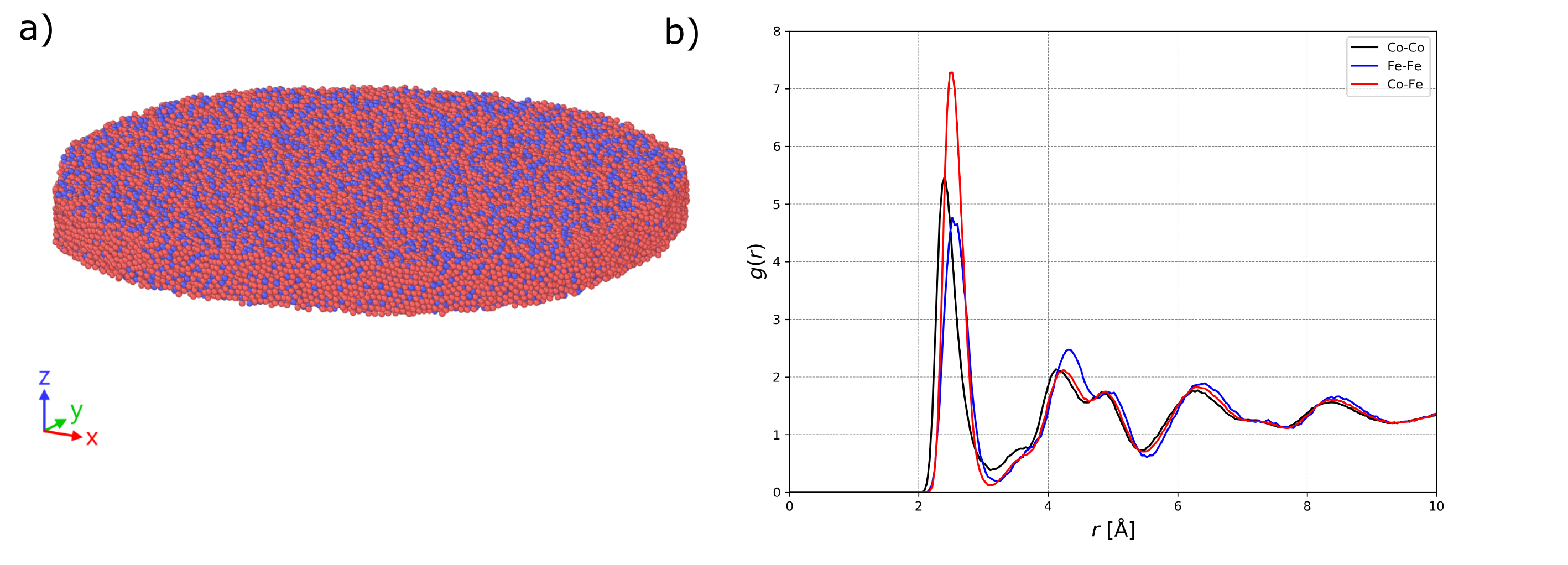}
\caption{Structural characterization of the amorphous $Co_{x}Fe_{1-x}$ nanodisk. (a) Atomistic 3D representation of the disk for the $Co_{65}Fe_{35}$ stoichiometry. (b) Partial radial distribution function, $g(r)$, where the sharp first peak and subsequent dampened oscillations confirm the stable amorphous phase.}\label{sample-gdr}
\end{figure}

Amorphous Co$_{x}$Fe$_{1-x}$ nanodisks with physical dimensions of $23.6 \times 23.6 \times 1.96$ nm$^3$ , shown in Fig.~\ref{sample-gdr}(a), were modeled across three stoichiometric ratios: Co$_{50}$Fe$_{50}$, Co$_{65}$Fe$_{35}$, and Co$_{35}$Fe$_{65}$. The initial configurations were derived from a B2-Cobalt crystalline structure with random Iron substitutions to match the target percentages. To achieve a representative amorphous state, a melt-quenching protocol proposed by Wang {\it et al.} was employed~\cite{Wang2012}. The systems were first heated to $2500$~K at $0$~GPa with a timestep of $\Delta t = 1$~fs, followed by cooling to $300$~K at a calculated rate of $10^{10}$~K/s. Finally, the samples were equilibrated in the NVT ensemble for $100$~ps.

Interatomic interactions were modeled using the modified embedded-atom method (MEAM) potential~\cite{Sharifi2025}. Structural analysis and visualization were performed using the OVITO software~\cite{Stukowski2010}. The amorphization was validated through the partial radial distribution function (RDF), $g(r)$, as shown in Fig.~\ref{sample-gdr}(b). For the representative Co$_{65}$Fe$_{35}$ stoichiometry, the RDF profile exhibits a prominent first-neighbor peak at $\approx 2.5$ \AA\ and a split second peak, consistent with short-range order in metallic glasses~\cite{Sheng2006}. The absence of long-range periodic peaks beyond $10$ \AA\ confirms the successful amorphization of the nanodisks. Local coordination was further evaluated via Voronoi polyhedra classification.

\subsection{Magnetic Calibration and Spin-Lattice Dynamics}
To construct the magnetic topology, $J_{ij}$ was defined as a distance-dependent function to account for the disordered lattice:

\begin{equation}
    J_{ij} = 4a \left( \frac{r_{ij}}{d} \right)^{2} \left( 1 - b \left( \frac{r_{ij}}{d} \right)^{2} \right) e^{-(r_{ij}/d)^{2}}
\end{equation}

where $\{a, b, d\}$ are constants defining the interaction range. The global scaling of $J_{ij}$ was performed by estimating the Curie temperature ($T_C$) using the Mean-Field Approximation (MFA) for classical Heisenberg spins~\cite{Evans2014}:

\begin{equation}
    T_{C}^{MF} = \frac{2 \overline{J_{0}}}{3 k_{B}}, \quad \text{with} \quad \overline{J_{0}} = \langle \sum_{j} J_{ij} \rangle
\end{equation}

The exchange constants were adjusted to align with experimental benchmarks for amorphous Co--Fe alloys, where $T_C$ is set to be $\approx 800$~K for Co$_{50}$Fe$_{50}$~\cite{Willard1998} to lower values in Iron-rich compositions~\cite{Son2023}. A global scaling factor $f = T_{C}^{\text{target}} / T_{C}^{\text{current}}$ was applied to ensure the modeled ferromagnetic-to-paramagnetic transition accounts for the inherent topological disorder.

Finally, spin-lattice dynamics (SLD) simulations were conducted for $100$~ps with a constant Zeeman field of $0.1$~T applied along the $z$-axis. This dual-timestep approach captures the feedback loop between structural vibrations and spin precession.

\section{Results and Discussion}

\subsection{Local Structural Heterogeneity in Amorphous Co$_{x}$Fe$_{1-x}$}
\begin{figure}[t]%% placement specifier
\centering%% For centre alignment of image.
\includegraphics[width=1.0\textwidth]{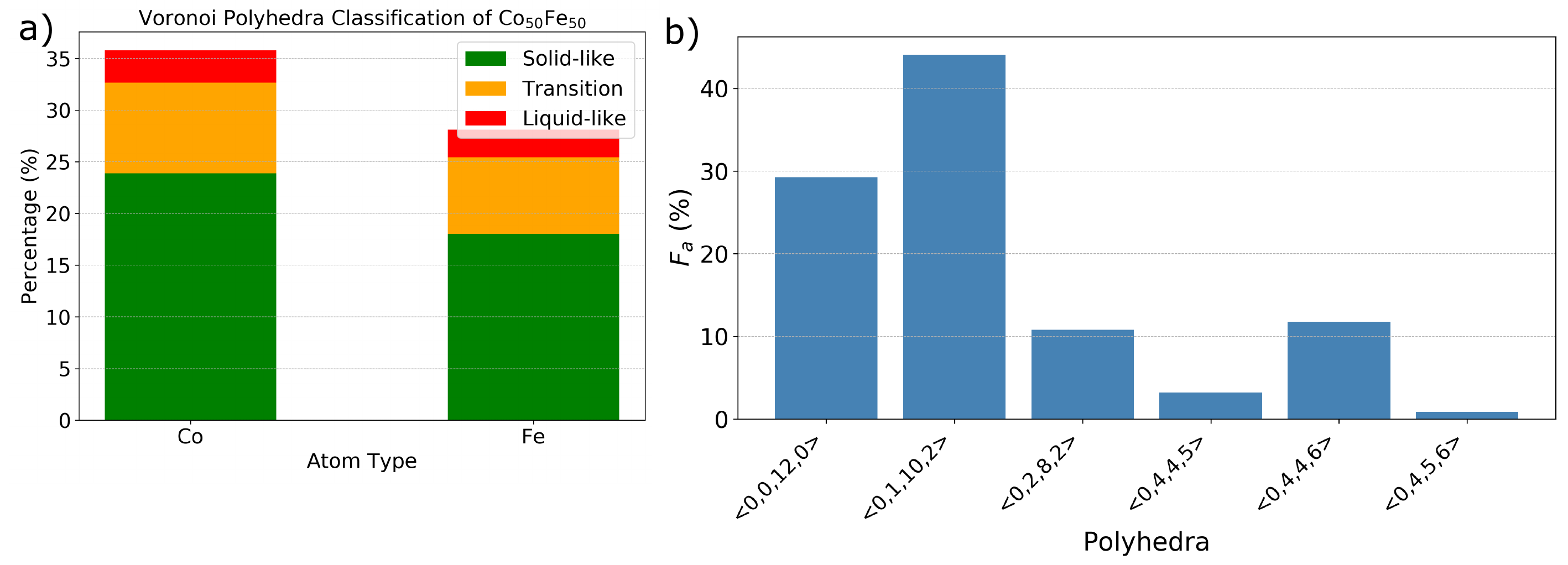}
%% Use \caption command for figure caption and label.
\caption{Topological analysis of the local atomic environment for the Co$_{50}$Fe$_{50}$ nanodisk. (a) Voronoi polyhedra classification by atomic species, showing the distribution of ``solid-like'', ``transition'', and ``liquid-like'' coordination environments. (b) Frequency distribution of specific Voronoi indices $\langle n_3, n_4, n_5, n_6 \rangle$, highlighting the dominance of icosahedral-like packing motifs such as $\langle 0, 1, 10, 2 \rangle$ and $\langle 0, 0, 12, 0 \rangle$.}\label{res5050}
\end{figure}

The local structural characteristics of the amorphous $Co_{x}Fe_{1-x}$ nanodisks were further elucidated through a detailed Voronoi tessellation analysis. Fig.~\ref{res5050}(a) presents the categorical classification of coordination polyhedra into ``solid-like'', ``transition'', and ``liquid-like'' environments for the Co$_{50}$Fe$_{50}$ stoichiometry. This species-dependent analysis reveals a significant structural disparity: while Cobalt atoms exhibit a robust dominance of ``solid-like'' configurations (approximately $26\%$), Iron atoms are characterized by a higher degree of local disorder, with a substantially lower solid-like fraction of $16\%$. This heterogeneity is further quantified in Fig.~\ref{res5050}(b), which displays the distribution of specific Voronoi indices $\langle n_3, n_4, n_5, n_6 \rangle$. The population is notably dominated by the $\langle 0, 1, 10, 2 \rangle$ and $\langle 0, 0, 12, 0 \rangle$ polyhedra, accounting for over $40\%$ and nearly $30\%$ of the local structures, respectively. These indices are indicative of high-coordination icosahedral-like packing, which serves as the primary structural motif for stabilizing the amorphous matrix against crystallization. The prevalence of these dense, solid-like clusters—particularly around Cobalt sites—provides the essential short-range order required to maintain a strong ferromagnetic exchange network while simultaneously facilitating the soft magnetic behavior inherent to the disordered nanostructure.

\begin{figure}[t]%% placement specifier
\centering%% For centre alignment of image.
\includegraphics[width=1.0\textwidth]{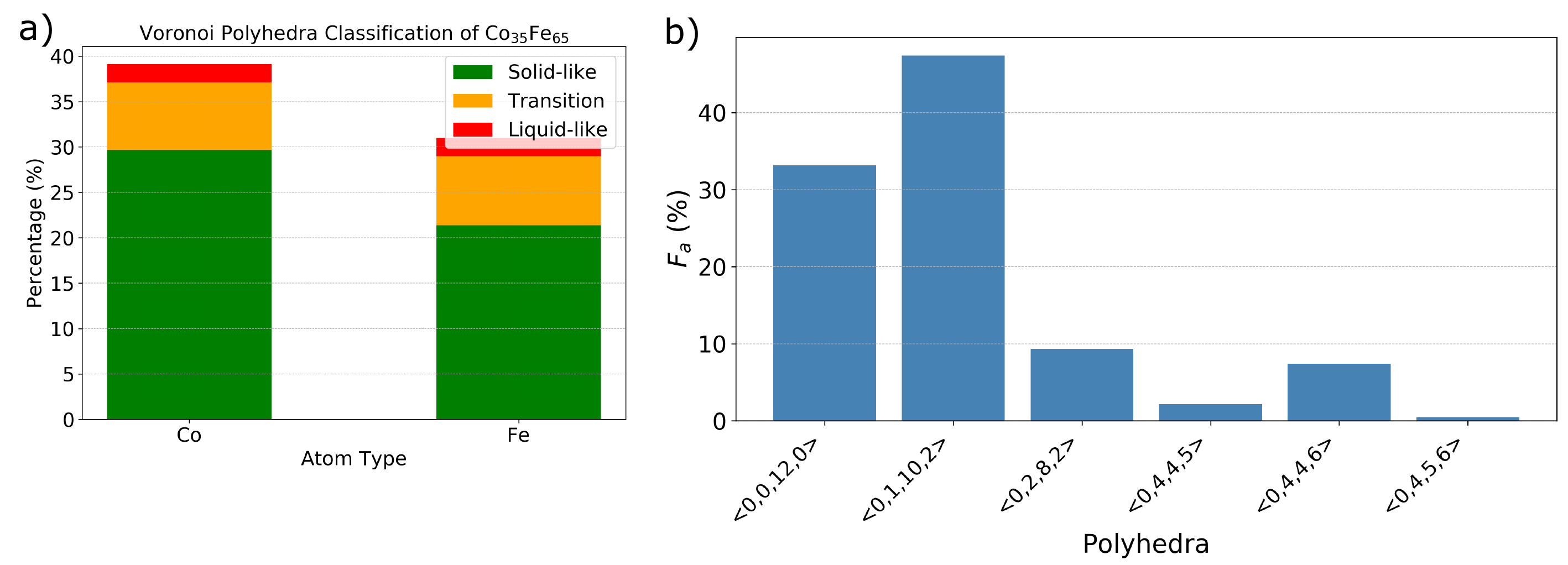}
%% Use \caption command for figure caption and label.
\caption{Structural characterization of the iron-rich Co$_{35}$Fe$_{65}$ stoichiometry. (a) Species-dependent Voronoi classification, illustrating the persistent higher degree of local order around Cobalt atoms compared to Iron. (b) Distribution of coordination polyhedra indices, indicating a high prevalence of dense packing units that stabilize the amorphous matrix despite the increased Iron concentration.}\label{res3565}
\end{figure}

The local coordination environment for the Co$_{35}$Fe$_{65}$ stoichiometry was further analyzed by calculating the distribution of specific Voronoi indices $\langle n_3, n_4, n_5, n_6 \rangle$, as shown in Fig.~\ref{res3565}(b). The topological analysis reveals that the amorphous matrix is predominantly composed of icosahedral-like structures, with the $\langle 0, 1, 10, 2 \rangle$ and $\langle 0, 0, 12, 0 \rangle$ polyhedra being the most frequent, accounting for approximately $48\%$ and $33\%$ of the total population, respectively. These high-coordination motifs are characteristic of the short-range order (SRO) that stabilizes the amorphous phase in Co--Fe alloys. The high frequency of these indices, particularly the perfect icosahedron $\langle 0, 0, 12, 0 \rangle$, confirms a dense atomic packing that correlates with the "solid-like" fraction identified in the previous species-dependent classification. In this Iron-rich composition, the prevalence of these stabilizing motifs around Cobalt atoms persists, providing a robust structural framework that maintains magnetic connectivity despite the increased global disorder introduced by the higher Iron content.

\begin{figure}[h!]
    \centering
    \includegraphics[width=1.0\textwidth]{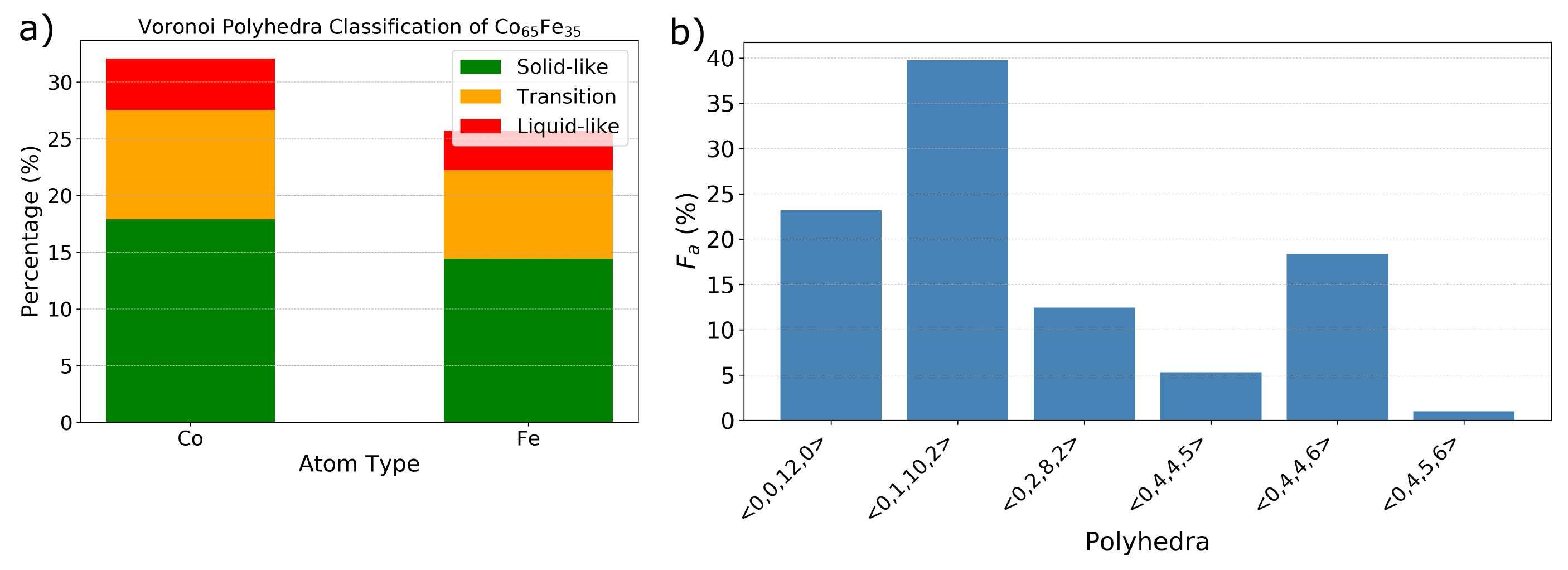}
    \caption{Structural characterization of the Co$_{65}$Fe$_{35}$ stoichiometry. (a) Voronoi polyhedra classification illustrating the dominance of transition and "solid-like" environments. (b) Frequency distribution of Voronoi indices, highlighting the prevalence of icosahedral packing motifs that stabilize the amorphous phase.}
    \label{res_6535}
\end{figure}

The local atomic environment of the Co$_{65}$Fe$_{35}$ nanodisk was characterized through Voronoi tessellation to evaluate the impact of high cobalt concentration on the amorphous stability. As shown in the classification results (Fig.~\ref{res_6535}a), Cobalt atoms exhibit a distinct distribution where ``transition'' polyhedra represent the most prominent fraction, exceeding $27\%$, followed by ``solid-like'' environments which account for approximately $18\%$ of the local coordination. In comparison, Iron atoms show a lower prevalence of ordered structures, confirming that the structural backbone of the alloy is primarily supported by the Cobalt phase.

This topological arrangement is further detailed in the distribution of Voronoi indices $\langle n_3, n_4, n_5, n_6 \rangle$ (Fig.~\ref{res_6535}b). The short-range order (SRO) is significantly dominated by icosahedral-like motifs, with the $\langle 0, 1, 10, 2 \rangle$ polyhedron being the most frequent, representing nearly $40\%$ of the total population. Additionally, the perfect icosahedron $\langle 0, 0, 12, 0 \rangle$ contributes approximately $23\%$ to the coordination shells. The high concentration of these dense, high-coordination units—particularly the dominance of the $\langle 0, 1, 10, 2 \rangle$ motif—indicates a stable and well-packed amorphous matrix, which is essential for the subsequent investigation of the magnetic exchange and spin dynamics in this Cobalt-rich stoichiometry.

A comprehensive comparison across the Co$_{50}$Fe$_{50}$, Co$_{35}$Fe$_{65}$, and Co$_{65}$Fe$_{35}$ stoichiometries reveals a consistent trend in the formation of short-range order within the amorphous Co-Fe system. In all studied cases, Cobalt atoms demonstrate a superior degree of local organization compared to Iron atoms, with the Co$_{65}$Fe$_{35}$ configuration reaching a combined solid-like and transition fraction that exceeds $45\%$ for the Cobalt species. This confirms that Cobalt consistently acts as the primary topological stabilizer, fostering the growth of dense atomic clusters regardless of the global chemical ratio.

The analysis of the Voronoi indices across the three systems highlights the universality of icosahedral packing in these metallic glasses. The persistence of the $\langle 0, 1, 10, 2 \rangle$ and $\langle 0, 0, 12, 0 \rangle$ motifs as the dominant structures, especially reaching their highest frequency in the Co$_{65}$Fe$_{35}$ sample, indicates a reduction in local free volume as Cobalt content increases. This topological consistency is a critical finding for the magnetic modeling of the alloy, as it provides a reproducible atomistic baseline. The stable structural motifs ensure that the distance-dependent exchange interaction parameters ($J_{ij}$) effectively capture the magnetic connectivity required to simulate accurate spin-lattice dynamics across the entire range of studied compositions.

\subsection{Impact of Local Topology on Magnetic Calibration}

The observed topological contrast between species has direct implications for the magnetic response of the Co--Fe nanodisks. The dominance of solid-like polyhedra around Cobalt atoms ensures a more stable and uniform exchange interaction environment ($J_{ij}$) in these regions. In contrast, the more disordered neighborhoods of the Iron atoms introduce local fluctuations in the magnetic coupling constants. 

This structural-magnetic interplay was explicitly accounted for during the $J_{ij}$ calibration process. By utilizing the structural \texttt{dump} files from the equilibrated MD samples, the distance-dependent exchange parameters were scaled to align the system's Mean-Field Curie temperature ($T_C^{MF}$) with experimental benchmarks. For the Co$_{50}$Fe$_{50}$ stoichiometry, the parameters were adjusted to reflect a $T_C$ of approximately $1120$ K, while Iron-rich environments in the Co$_{35}$Fe$_{65}$ case exhibited lower estimated transition temperatures due to the increased prevalence of disordered ``liquid-like'' polyhedra. This multiscale structural arrangement ensures that the calibrated exchange landscape accurately reflects the intrinsic damping and spin-wave propagation modes within the amorphous nanostructure.

\subsection{Magnetization Dynamics and Relaxation}

\begin{figure}[h!]
    \centering
    \includegraphics[width=0.65\textwidth]{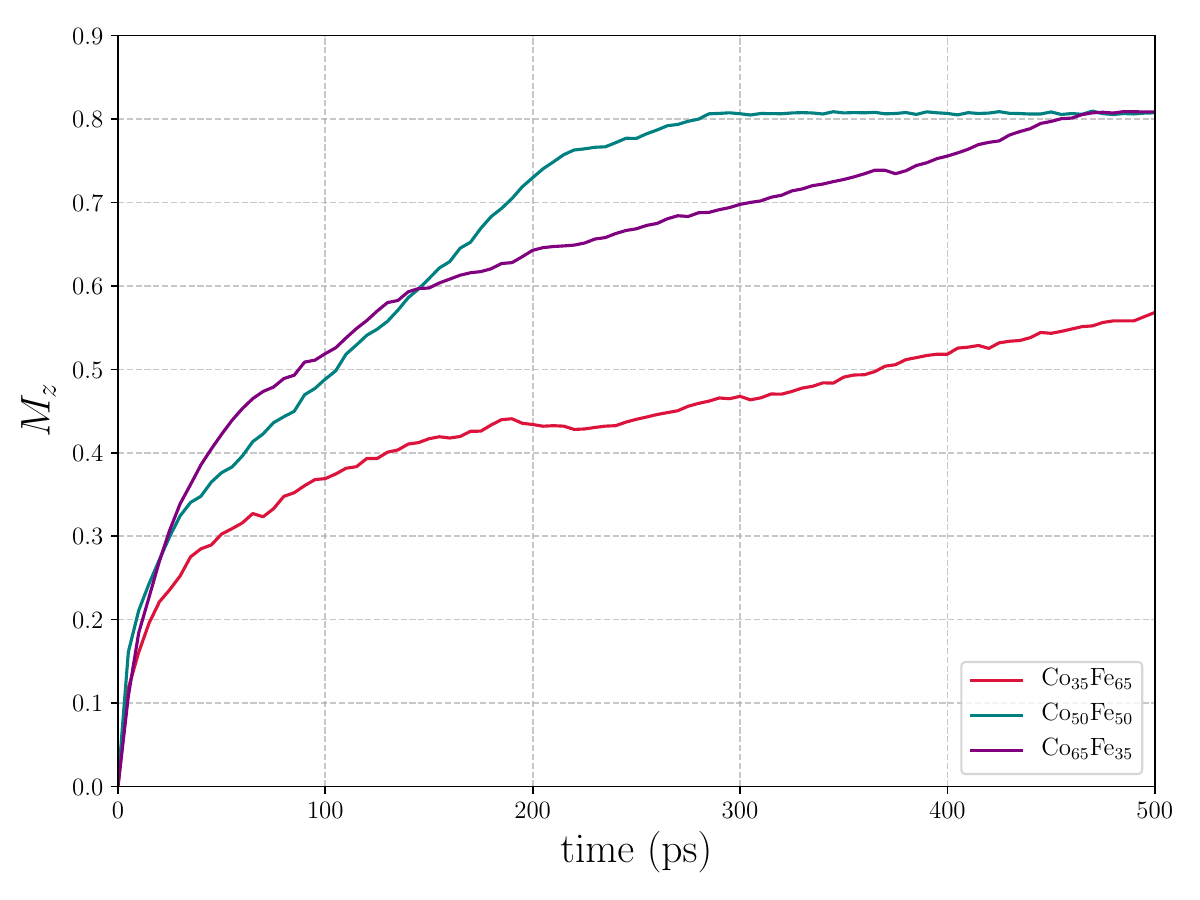}
    \caption{Longitudinal magnetization relaxation $M_z$ as a function of time for Co$_{35}$Fe$_{65}$, Co$_{50}$Fe$_{50}$, and Co$_{65}$Fe$_{35}$ amorphous nanodisks. The curves demonstrate the impact of chemical composition on the saturation magnetization and relaxation kinetics.}
    \label{fig:mz_relaxation}
\end{figure}

The temporal evolution of the longitudinal magnetization, $M_z(t)$, for the three studied amorphous Co$_{x}$Fe$_{1-x}$ nanodisks is illustrated in Fig.~\ref{fig:mz_relaxation}. The relaxation process under an external magnetic field of $0.1$ T reveals a clear dependence on the stoichiometric ratio, where the rate of alignment and the saturation levels are intrinsically linked to the chemical and structural disorder of the samples. All compositions exhibit a characteristic ferromagnetic-like relaxation, yet the dynamic response varies significantly as the Cobalt content increases.

The Co$_{65}$Fe$_{35}$ stoichiometry displays the most efficient magnetic relaxation, characterized by the steepest initial slope and the highest saturation magnetization, reaching $M_z \approx 0.81$ within the first $500$ ps. This enhanced performance is directly correlated with the local structural findings, where Cobalt atoms were found to maintain a superior degree of "solid-like" coordination and a higher prevalence of dense icosahedral motifs compared to Iron. This robust structural backbone minimizes the fluctuations in the exchange integrals ($J_{ij}$), facilitating a more coherent and rapid alignment of the magnetic moments. In contrast, the iron-rich Co$_{35}$Fe$_{65}$ sample exhibits the slowest relaxation and the lowest saturation plateau ($M_z \approx 0.57$). The increased presence of Iron-rich environments, which are characterized by a higher fraction of "liquid-like" and transition polyhedra, introduces significant topological disorder that disrupts the exchange connectivity and enhances the magnetic damping within the nanodisk.

The equiatomic Co$_{50}$Fe$_{50}$ configuration represents an intermediate state, reaching a stable saturation of $M_z \approx 0.80$ at approximately $300$ ps. Interestingly, while its saturation level is comparable to that of the Co$_{65}$Fe$_{35}$ sample, its initial relaxation rate is slightly lower, reflecting the balanced influence of both species on the global exchange field. These results confirm that the macroscopic magnetic relaxation in $Co-Fe$ amorphous alloys is fundamentally governed by the short-range order, where the stabilization of dense packing units provided by Cobalt is essential for maintaining high saturation and fast dynamic response.

The correlation between the icosahedral short-range order (SRO) and the magnetic stability in Co$_{65}$Fe$_{35}$ aligns with recent findings in transition-metal amorphous alloys. According to Gómez-Polo et al. \cite{GomezPolo1996}, the soft magnetic properties in amorphous systems are governed by the distribution of local exchange interactions. In our Co-Fe system, we identify that these interactions are stabilized by high-coordination units such as the $\langle 0, 1, 10, 2 \rangle$ motifs identified in our work. Furthermore, the higher saturation and faster relaxation observed in Co$_{65}$Fe$_{35}$ compared to Co$_{35}$Fe$_{65}$ are consistent with reports by Son {\it et al.}, who demonstrated that compositions favoring dense atomic networks minimize magnetic damping and optimize magnetization pathways~\cite{Son2023}. These results confirm that the transition from disordered iron-rich environments to cobalt-stabilized icosahedral clusters is the primary descriptor for the dynamic response in these nanodisks.

\section{Conclusions}

In this work, we have investigated the structural and magnetic properties of amorphous Co$_{x}$Fe$_{1-x}$ nanodisks using a combined molecular dynamics and atomistic spin dynamics approach. Our results demonstrate that the local atomic arrangement, characterized through Voronoi tessellation, plays a fundamental role in determining the macroscopic magnetic response of these metallic glasses. Across all studied stoichiometries, we identified a persistent structural heterogeneity where Cobalt atoms consistently promote a higher degree of short-range order compared to Iron atoms. Specifically, the Co$_{65}$Fe$_{35}$ and Co$_{50}$Fe$_{50}$ configurations exhibit a robust framework of ``solid-like'' and ``transition'' polyhedra, dominated by icosahedral packing motifs such as the $\langle 0, 1, 10, 2 \rangle$ and $\langle 0, 0, 12, 0 \rangle$ indices.

The dynamic simulations reveal that this structural stability is intrinsically linked to the magnetic relaxation process. The Co$_{65}$Fe$_{35}$ stoichiometry, characterized by the highest fraction of dense packing units, displays the most efficient magnetic alignment and the highest saturation magnetization ($M_z \approx 0.81$). In contrast, the increase in Iron content in the Co$_{35}$Fe$_{65}$ sample leads to a higher prevalence of disordered ``liquid-like'' environments, which enhances magnetic damping and results in a significantly lower saturation plateau ($M_z \approx 0.57$). These findings suggest that Cobalt acts as a topological stabilizer that maintains the exchange connectivity ($J_{ij}$) necessary for rapid spin alignment. This research provides insights into the design of amorphous magnetic nanomaterials, highlighting that precise control over the local coordination environment is essential for optimizing the performance of future magnonic and spintronic devices.

%Such structural heterogeneity at the local scale, particularly the dominance of solid-like Voronoi polyhedra around Cobalt atoms compared to the more disordered environments of Iron atoms, plays a decisive role in the resulting magnetic damping and spin-wave propagation within these nanostructures.

\bibliographystyle{IEEEtran}

\addcontentsline{toc}{chapter}{Bibliography}
\bibliography{biblio}

@article{McHenry1999,
title = {Amorphous and nanocrystalline materials for applications as soft magnets},
journal = {Progress in Materials Science},
volume = {44},
number = {4},
pages = {291-433},
year = {1999},
issn = {0079-6425},
doi = {https://doi.org/10.1016/S0079-6425(99)00002-X},
url = {https://www.sciencedirect.com/science/article/pii/S007964259900002X},
author = {Michael E McHenry and Matthew A Willard and David E Laughlin}
}

@incollection{Lubrorsky1980,
title = {Chapter 6 Amorphous ferromagnets},
series = {Handbook of Ferromagnetic Materials},
publisher = {Elsevier},
volume = {1},
pages = {451-529},
year = {1980},
issn = {1574-9304},
doi = {https://doi.org/10.1016/S1574-9304(05)80121-X},
url = {https://www.sciencedirect.com/science/article/pii/S157493040580121X},
author = {F.E. Luborsky}
}

@article{Bai2018,
AUTHOR = {Yang, Bai and Wu, Yue and Li, Xiaopan and Yu, Ronghai},
TITLE = {Chemical Synthesis of High-Stable Amorphous FeCo Nanoalloys with Good Magnetic Properties},
JOURNAL = {Nanomaterials},
VOLUME = {8},
YEAR = {2018},
NUMBER = {3},
ARTICLE-NUMBER = {154},
URL = {https://www.mdpi.com/2079-4991/8/3/154},
PubMedID = {29522436},
ISSN = {2079-4991},
DOI = {10.3390/nano8030154}
}

@article{Miracle2004,
  title = {A structural model for metallic glasses},
  author = {Miracle, Daniel B},
  journal = {Nature Materials},
  volume = {3},
  pages = {697--702},
  year = {2004},
  publisher = {Nature Publishing Group},
url={https://www.nature.com/articles/nmat1219},
  doi = {10.1038/nmat1219}

}

@article{KAUL19855,
title = {Static critical phenomena in ferromagnets with quenched disorder},
journal = {Journal of Magnetism and Magnetic Materials},
volume = {53},
number = {1},
pages = {5-53},
year = {1985},
issn = {0304-8853},
doi = {https://doi.org/10.1016/0304-8853(85)90128-3},
url = {https://www.sciencedirect.com/science/article/pii/0304885385901283},
author = {S.N. Kaul}
}

@article{Tsvelick01011983,
author = {A.M. Tsvelick and P.B. Wiegmann},
title = {Exact results in the theory of magnetic alloys},
journal = {Advances in Physics},
volume = {32},
number = {4},
pages = {453--713},
year = {1983},
publisher = {Taylor \& Francis},
doi = {10.1080/00018738300101581},
URL = {https://doi.org/10.1080/00018738300101581},
eprint = { https://doi.org/10.1080/00018738300101581}
}

@article{BHATTI2017530,
title = {Spintronics based random access memory: a review},
journal = {Materials Today},
volume = {20},
number = {9},
pages = {530-548},
year = {2017},
issn = {1369-7021},
doi = {https://doi.org/10.1016/j.mattod.2017.07.007},
url = {https://www.sciencedirect.com/science/article/pii/S1369702117304285},
author = {Sabpreet Bhatti and Rachid Sbiaa and Atsufumi Hirohata and Hideo Ohno and Shunsuke Fukami and S.N. Piramanayagam}
}

@article{Ma2011,
  title = {Langevin spin dynamics},
  author = {Ma, Pui Wai and Dudarev, S. L.},
  journal = {Phys. Rev. B},
  volume = {83},
  issue = {13},
  pages = {134418},
  numpages = {9},
  year = {2011},
  month = {Apr},
  publisher = {American Physical Society},
  doi = {10.1103/PhysRevB.83.134418},
  url = {https://link.aps.org/doi/10.1103/PhysRevB.83.134418}
}

@article{Ma2012,
  title = {Longitudinal magnetic fluctuations in Langevin spin dynamics},
  author = {Ma, Pui Wai and Dudarev, S. L.},
  journal = {Phys. Rev. B},
  volume = {86},
  issue = {5},
  pages = {054416},
  numpages = {10},
  year = {2012},
  month = {Aug},
  publisher = {American Physical Society},
  doi = {10.1103/PhysRevB.86.054416},
  url = {https://link.aps.org/doi/10.1103/PhysRevB.86.054416}
}

@article{Beaujouan2012,
   title = {Anisotropic magnetic molecular dynamics of cobalt nanowires},
  author = {Beaujouan, David and Thibaudeau, Pascal and Barreteau, Cyrille},
  journal = {Phys. Rev. B},
  volume = {86},
  issue = {17},
  pages = {174409},
  numpages = {11},
  year = {2012},
  month = {Nov},
  publisher = {American Physical Society},
  doi = {10.1103/PhysRevB.86.174409},
  url = {https://link.aps.org/doi/10.1103/PhysRevB.86.174409}
}

@article{Tranchida2018,
 title = {Massively parallel symplectic algorithm for coupled magnetic spin dynamics and molecular dynamics},
journal = {Journal of Computational Physics},
volume = {372},
pages = {406-425},
year = {2018},
issn = {0021-9991},
doi = {https://doi.org/10.1016/j.jcp.2018.06.042},
url = {https://www.sciencedirect.com/science/article/pii/S0021999118304200},
author = {J. Tranchida and S.J. Plimpton and P. Thibaudeau and A.P. Thompson}
}

@article{Oogane2006,
doi = {10.1143/JJAP.45.3889},
url = {https://doi.org/10.1143/JJAP.45.3889},
year = {2006},
month = {may},
publisher = {},
volume = {45},
number = {5R},
pages = {3889},
author = {Oogane, Mikihiko and Wakitani, Takeshi and Yakata, Satoshi and Yilgin, Resul and Ando, Yasuo and Sakuma, Akimasa and Miyazaki, Terunobu},
title = {Magnetic Damping in Ferromagnetic Thin Films},
journal = {Japanese Journal of Applied Physics}
}

@article{lammps,
  author = "A. P. Thompson and H. M. Aktulga and R. Berger and 
     D. S. Bolintineanu and W. M. Brown and P. S. Crozier and
     P. J. in 't Veld and A. Kohlmeyer and S. G. Moore and T. D. Nguyen and
     R. Shan and M. J. Stevens and J. Tranchida and C. Trott and S. J. Plimpton",
  title = "{LAMMPS} - a flexible simulation tool for
     particle-based materials modeling at the 
     atomic, meso, and continuum scales",
  journal = "Comp. Phys. Comm.",
  volume =  "271",
  pages =   "108171",
  year =    "2022",
  doi = "10.1016/j.cpc.2021.108171"
}

@article{Wang2012,
title = {Structural properties of Zr$_{x}$Cu$_{90-x}$Al$_{10}$ metallic glasses investigated by molecular dynamics simulations},
journal = {Journal of Alloys and Compounds},
volume = {510},
number = {1},
pages = {107-113},
year = {2012},
issn = {0925-8388},
doi = {https://doi.org/10.1016/j.jallcom.2011.07.110},
url = {https://www.sciencedirect.com/science/article/pii/S0925838811017816},
author = {C.C. Wang and C.H. Wong},
}

@article{Stukowski2010,
doi = {10.1088/0965-0393/18/1/015012},
url = {https://doi.org/10.1088/0965-0393/18/1/015012},
year = {2009},
month = {dec},
publisher = {},
volume = {18},
number = {1},
pages = {015012},
author = {Stukowski, Alexander},
title = {Visualization and analysis of atomistic simulation data with OVITO–the Open Visualization Tool},
journal = {Modelling and Simulation in Materials Science and Engineering},
}

@article{Sharifi2025,
title = {Developing interatomic potentials for complex concentrated alloys of Cu, Ti, Ni, Cr, Co, Al, Fe, and Mn},
journal = {Computational Materials Science},
volume = {248},
pages = {113595},
year = {2025},
issn = {0927-0256},
doi = {https://doi.org/10.1016/j.commatsci.2024.113595},
url = {https://www.sciencedirect.com/science/article/pii/S0927025624008164},
author = {Hamid Sharifi and Collin D. Wick},
}

@article{Evans2014,
doi = {10.1088/0953-8984/26/10/103202},
url = {https://doi.org/10.1088/0953-8984/26/10/103202},
year = {2014},
month = {feb},
publisher = {IOP Publishing},
volume = {26},
number = {10},
pages = {103202},
author = {Evans, R F L and Fan, W J and Chureemart, P and Ostler, T A and Ellis, M O A and Chantrell, R W},
title = {Atomistic spin model simulations of magnetic nanomaterials},
journal = {Journal of Physics: Condensed Matter},
}

@article{Sheng2006,
  author  = {Sheng, H. W. and Luo, W. K. and Alamgir, F. M. and Bai, J. M. and Ma, E.},
  title   = {Atomic packing and short-to-medium-range order in metallic glasses},
  journal = {Nature},
  year    = {2006},
  volume  = {439},
  number  = {7075},
  pages   = {419--425},
  month   = {jan},
  doi     = {10.1038/nature04421},
  url     = {https://doi.org/10.1038/nature04421},
  issn    = {1476-4687}
}

@Article{Son2023,
AUTHOR = {Son, Hyunsol and Park, Jihye and Lee, Hyunkyung and Choi-Yim, Haein},
TITLE = {Annealing Effect in Amorphous Fe-Co-B-Si-Nb According to Fe/Co Ratio},
JOURNAL = {Metals},
VOLUME = {13},
YEAR = {2023},
NUMBER = {4},
ARTICLE-NUMBER = {715},
URL = {https://www.mdpi.com/2075-4701/13/4/715},
ISSN = {2075-4701},
DOI = {10.3390/met13040715}
}

@article{Willard1998,  author = {Willard, M. A. and Laughlin, D. E. and McHenry, M. E. and Thoma, D. and Sickafus, K. and Cross, J. O. and Harris, V. G.},
    title = {Structure and magnetic properties of (Fe$_{0.5}$Co$_{0.5}$)$_{88}$Zr$_{7}$B$_{4}$Cu$_{1}$ nanocrystalline alloys},
    journal = {Journal of Applied Physics},
    volume = {84},
    number = {12},
    pages = {6773-6777},
    year = {1998},
    month = {12},
    issn = {0021-8979},
    doi = {10.1063/1.369007},
    url = {https://doi.org/10.1063/1.369007},
    eprint = {https://pubs.aip.org/aip/jap/article-pdf/84/12/6773/19121036/6773_1_online.pdf},
}

@article{GomezPolo1996,
  title = {Distribution of exchange fields in amorphous ferromagnets},
  author = {Gómez-Polo, C. and Vázquez, M. and Hernando, A.},
  journal = {Physical Review B},
  volume = {53},
  number = {6},
  pages = {3392--3400},
  year = {1996},
  publisher = {American Physical Society},
  doi = {10.1103/PhysRevB.53.3392},
url={https://journals.aps.org/prb/abstract/10.1103/PhysRevB.53.3392}
}

@article{Egami1984,
title = {Atomic size effect on the formability of metallic glasses},
journal = {Journal of Non-Crystalline Solids},
volume = {64},
number = {1},
pages = {113-134},
year = {1984},
issn = {0022-3093},
doi = {https://doi.org/10.1016/0022-3093(84)90210-2},
url = {https://www.sciencedirect.com/science/article/pii/0022309384902102},
author = {T Egami and Y Waseda}
}

@article{Cheng2011,
title = {Atomic-level structure and structure–property relationship in metallic glasses},
journal = {Progress in Materials Science},
volume = {56},
number = {4},
pages = {379-473},
year = {2011},
issn = {0079-6425},
doi = {https://doi.org/10.1016/j.pmatsci.2010.12.002},
url = {https://www.sciencedirect.com/science/article/pii/S0079642510000691},
author = {Y.Q. Cheng and E. Ma}
}

\end{document}